\documentclass[bibyear]{aa} 

\usepackage{graphicx}
\usepackage{amsmath}
\usepackage{float} 

\usepackage{txfonts}
\begin{document}

  \title{A new method to compute limb-darkening coefficients for stellar atmosphere models with spherical symmetry:   the space missions {{\sc TESS}}, {{\sc Kepler}}, {{\sc Corot}}, and 
        {{\sc MOST}}  
 }
 { }
   \subtitle{  }
\author{A. Claret}
   \offprints{A. Claret, e-mail:claret@iaa.es. Tables 2-17 are   available 
in electronic form at the CDS via anonymous ftp. Additional  calculations for other photometric systems and/or other bi-parametric laws can be performed on request. }
\institute{Instituto de Astrof\'{\i}sica de Andaluc\'{\i}a, CSIC, Apartado 3004,
            18080 Granada, Spain}
            \date{Received; accepted; }
\abstract
   {}
   {One of the biggest problems we can encounter while dealing with the limb-darkening coefficients for stellar atmospheric models with spherical symmetry is the difficulty of adjusting both the limb and the central parts simultaneously. In particular, the regions near the drop-offs are not well reproduced for most models, depending on Teff, log g, or wavelength. Even if the law with four terms is used, these disagreements still persist. Here we introduce a new method that considerably improves the description of both the limb and the central parts and that will allow users to test models of stellar atmospheres with spherical symmetry more accurately in environments  such as exoplanetary transits, eclipsing binaries, etc.}
   {The method introduced here is simple. Instead of considering all the $\mu $ points in the adjustment, as is traditional, we consider only the points until the drop-off ($\mu_{cri}$) of each model. From this point, we impose a condition ${I(\mu)}/{I (1)} = 0$. All calculations were performed by  adopting the least-squares method.} 
   { The resulting coefficients using this new method reproduce the intensity distribution of the PHOENIX spherical models (COND and DRIFT) quite well for the photometric systems of the space missions {{\sc TESS}}, {{\sc KEPLER}}, {{\sc COROT}}, and {{\sc MOST}}. The calculations cover the following ranges of local gravity and effective temperatures:  
        2.5 $\leq$  log g $\leq$ 6.0 and 1500 K $\leq$ T$_{\rm eff}$ 
        $\leq$ 12000 K.   The  new spherical coefficients can easily be adapted to the most commonly used light curve synthesis codes.  
 }
   {}

   \titlerunning {Limb-darkening coefficients }
   \maketitle
%

\section{Introduction}

The difficulty of finding good numerical adjustments simultaneously for the angular distribution of specific intensities of spherical stellar atmospheres models to derive  limb-darkening coefficients (LDC) is well known. To control this problem, some time ago some authors introduced the concept of quasi-spherical atmosphere. This concept allowed the use of spherical models, but without considering the regions near the limb (see e.g. Claret 2017). Another version of the quasi-spherical model was suggested by Wittkowskii,
Aufdenberg, \& Kervella (2004): instead of truncating the models at a certain value of  $\mu$ (often 0.10) the truncation was defined by searching for the maximum of the derivative 
of the specific intensity with respect to $r$ and shifting the profile to this point, where  $r = \sqrt{(1-\mu^2)}$ and $\mu$ is given by 
$\cos(\gamma)$, where $\gamma$ is the angle between the line of sight and  the outward surface normal. The resulting LDC reproduced the model intensities well, especially  when a four-term law was adopted. 

However,  if we consider all values of  $\mu$, i.e. including the drop-off regions, not even the law with  four terms was able to reproduce well the angular distribution of the specific intensities, except for a few values of  T$_{\rm eff}$ and log g. 
In this short paper, we introduce a new method that has proven  able to reproduce well the angular distribution of the specific intensities of spherical models (including the region of the drop-offs) for the first time. The accuracy of the adjustments does not depend on the effective temperatures or log g. 
The paper is organized as follows.  Section 2 introduces the new numerical method and the  methodology. The results of the calculations are presented and discussed in Section 3.

\section {Spherical symmetric LDC: a new numerical method}

As commented in the Introduction, it has been difficult to reconcile 
good  adjustments for the regions near the limb and for 
the central parts of spherical stellar atmosphere simultaneously. In this paper we introduce a simple but very effective method which is capable of describing with  high precision the behaviour of specific intensity distribution of stellar model atmospheres  with spherical symmetry (hereafter full spherical method,  FSM). The resulting improvements are particularly important in the regions located near the limb.  Often the traditional methods used to adjust spherical models considered all values of  $\mu$ in the fitting process, except those adopting the quasi-spherical or $r$-method procedures, but by definition  ignoring the regions near  the limb (see Claret 2017 and references  therein for a detailed discussion).  Here, instead, we consider for the fitting process only the points until the drop-off (hereafter $\mu_{cri}$) of each model.   The value of $\mu_{cri}$ for each model was obtained   by searching for the maximum of the derivative of
the specific intensity with respect to $r$. From this point, we impose the condition ${I(\mu)}/{I (1)} \equiv 0.0$. To guarantee a very precise fitting we  describe the selected part of the disk the law introduced by Claret (2000)

\begin{eqnarray}
\frac{I(\mu)}{ I(1)} = 1 - \sum_{k=1}^{4} {a_k} (1 - \mu^{\frac{k}{2}}),
\end{eqnarray}

\noindent
where $I(1)$ is the specific intensity at the centre of the disk and 
 $a_k$ are the corresponding LDC. We  also carried out adjustments for the quadratic law, which is the most frequently used bi-parametric law 
 
 \begin{eqnarray}
 \frac{I(\mu)}{ I(1)} = 1 - a(1 - \mu) - b(1 - \mu)^2, 
 \end{eqnarray}
 
\noindent
where $a$ and $b$ are the corresponding quadratic LDC.  
 The following models were used in this paper:  {\sc PHOENIX-COND}   (Husser et al. 2013), and {\sc PHOENIX-DRIFT}  (Witte et al. 2009), both with  spherical geometry (see Table A2).  The {\sc PHOENIX-DRIFT} models  are suitable for very cold configurations, such as late-M, L, and T-type dwarfs since they were generated considering a detailed computation of high-temperature condensed clouds. These grids cover, together with solar abundances, the following ranges of local gravity and effective temperatures:  
2.5 $\leq$  log g $\leq$ 6.0 and 1500 K $\leq$ T$_{\rm eff}$ 
$\leq$ 12000 K.  The model atmosphere specifc intensities were convolved with  the  transmission curves of the four mentioned space missions, and all calculations were performed by  adopting the least-squares method (LSM).

\begin{figure}
        \includegraphics[height=8.cm,width=6.4cm,angle=-90]{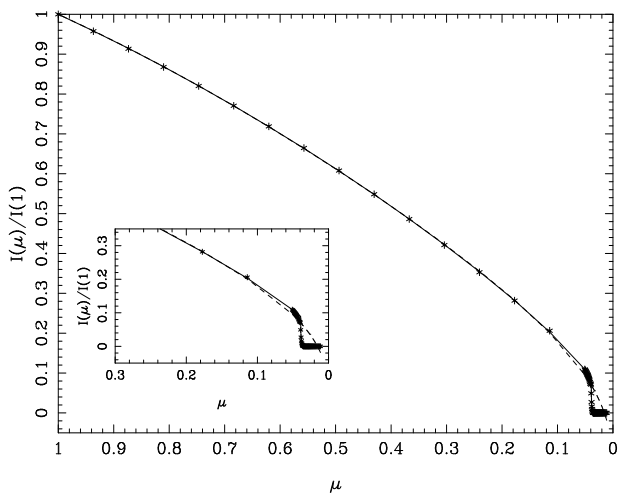}
        \caption{Specific intensity distribution for  a [T$_{\rm eff}$=2600 K, log g = 5.0] {\sc PHOENIX-DRIFT} 
                spherical symmetric  model. Asterisks represent the model intensities, 
                while the dashed line denotes the traditional fitting with four terms    
                and the continuous line  the fitting using the FSM: Eq. 1;  
                Log[A/H]= 0.0; $V_{\xi}$ = 2 km/s;  {\sc TESS} photometric system;  LSM calculations.}
\end{figure}

\begin{figure}
        \includegraphics[height=8.cm,width=6.4cm,angle=-90]{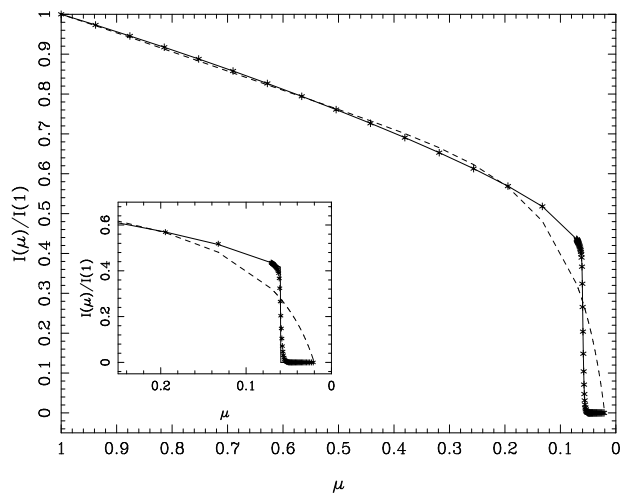}
        \caption{As in Fig. 1, but for a [T$_{\rm eff}$=5600 K, log g=4.5] {\sc PHOENIX-COND} 
                spherical symmetric  model.}
\end{figure}

\begin{figure}
        \includegraphics[height=8.cm,width=6.4cm,angle=-90]{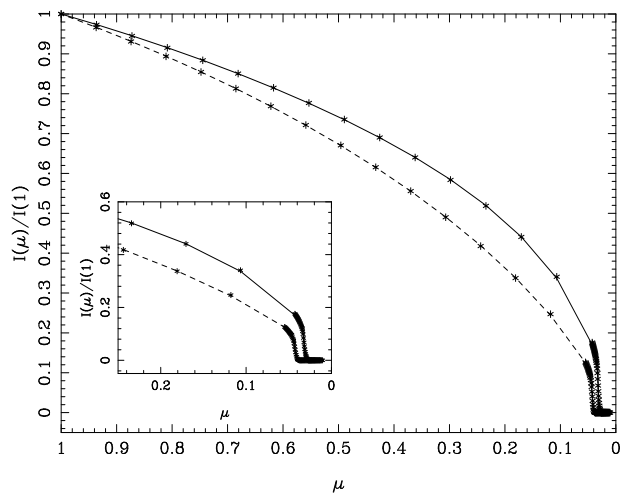}
        \caption{Specific intensity distribution for spherical symmetric models. The continuous line represents the   [T$_{\rm eff}$=2800 K, log g=5.0] {\sc PHOENIX-COND} model and the dashed line denotes the  corresponding  [T$_{\rm eff}$=2800 K, log g=5.0] {\sc PHOENIX-DRIFT}. Asterisks represent the model intensities.  Both fittings were performed with the FSM. Eq. 1;   
                Log[A/H]= 0.0; $V_{\xi}$ = 2 km/s;  {\sc TESS} photometric system;  LSM calculations.    }
\end{figure}

\begin{figure}
        \includegraphics[height=8.cm,width=6.4cm,angle=-90]{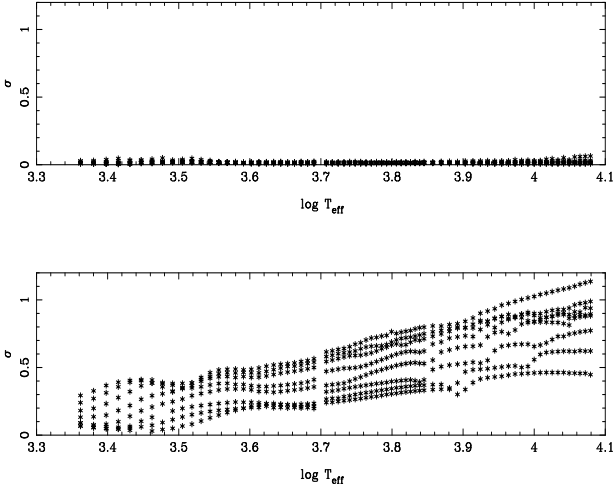}
        \caption{Square  root of the merit function adopting the FSM (first panel) and adopting the traditional method with four terms (second panel).  Only the PHOENIX-COND models listed in Table A2 are displayed. To guarantee the homogeneity of the comparison the results of the DRIFT models,  which were generated adopting a  different input physics, are not displayed.  
                Log[A/H]= 0.0; $V_{\xi}$ = 2 km/s; {\sc TESS} photometric system;  LSM calculations. }
\end{figure}

\begin{figure}
        \includegraphics[height=8.cm,width=6.4cm,angle=-90]{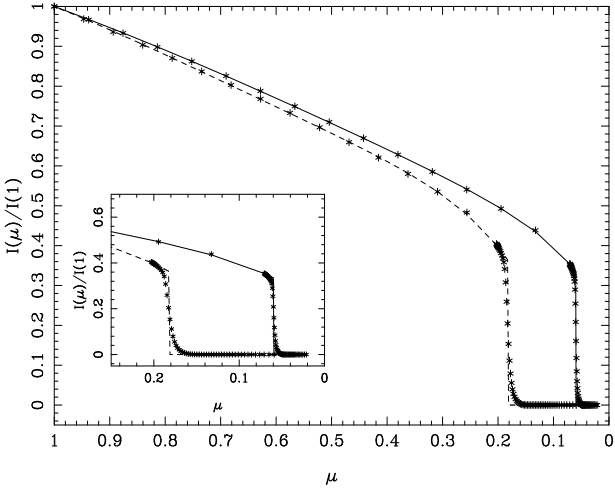}
        \caption{Effects of the local gravity on the specific intensity distribution. The continuous line represents the  [T$_{\rm eff}$=5600 K, log g=4.5] model and the dashed line   [T$_{\rm eff}$=5600 K, log g=2.5].  Asterisks represent the model intensities.  Both fittings were performed with the FSM. Eq. 1;    {\sc PHOENIX-COND} spherical symmetric models;  
                Log[A/H]= 0.0; $V_{\xi}$ = 2 km/s; {\sc KEPLER} photometric system;  LSM calculations. }
\end{figure}

\begin{figure}
        \includegraphics[height=8.cm,width=6.4cm,angle=-90]{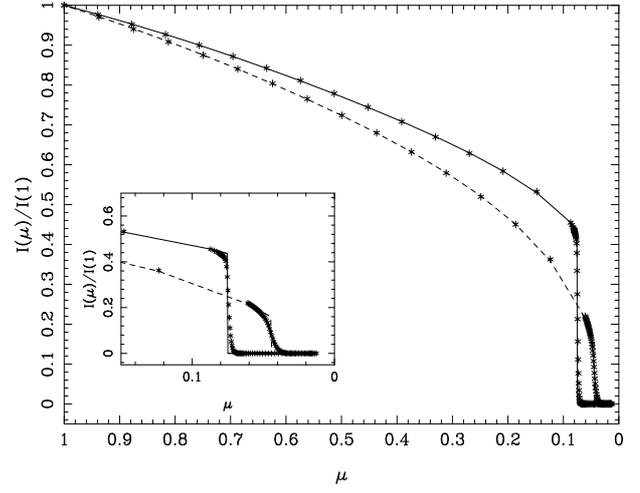}
        \caption{Effects of the effective temperature on the specific intensity distribution. The continuous line represents the  [T$_{\rm eff}$=8000 K, log g=4.5] model and the dashed line  [T$_{\rm eff}$=3000 K, log g=4.5]. Asterisks represent the model intensities. Both fittings were performed with the FSM.   {\sc PHOENIX-COND} spherical symmetric models;  Eq. 1;    
                Log[A/H]= 0.0; $V_{\xi}$ = 2 km/s; {\sc KEPLER} photometric system; LSM calculations. }
\end{figure}

\begin{figure}
        \includegraphics[height=8.cm,width=6.4cm,angle=-90]{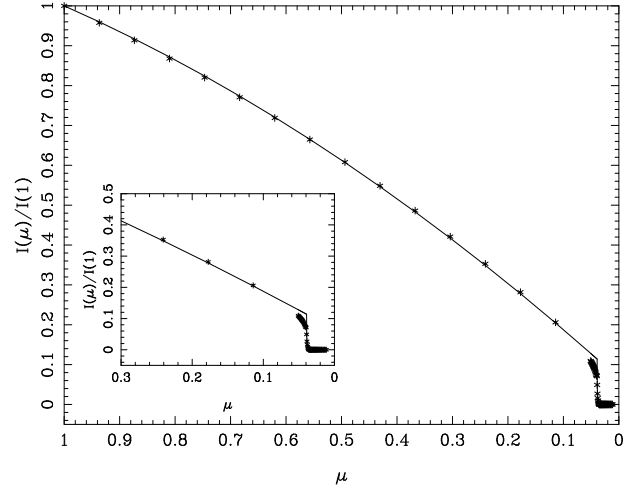}
        \caption{Specific intensity distribution for a   [T$_{\rm eff}$=2600 K, log g=5.0] {\sc PHOENIX-DRIFT} 
                spherical symmetric  model. Asterisks represent the model intensities, 
                while the continuous line represents the fitting using the FSM but adopting  Eq. 2;  
                Log[A/H]= 0.0; $V_{\xi}$ = 2 km/s; {\sc TESS} photometric system; LSM calculations.    }
\end{figure}

\section {Results and final remarks}

\renewcommand{\tablename}{ Table A}
\begin{table}
        \caption[]{Excerpt  from Table 7}
        \begin{flushleft}
                \begin{tabular}{ccccccl}                         
                        \hline                     
log g &Teff&  log Z&  L/HP&  a$_1$(KEPLER)\\
\hline
log g &Teff&  log Z&  L/HP&  a$_2$(KEPLER)\\
\hline
log g &Teff&  log Z&  L/HP&  a$_3$(KEPLER)\\
\hline
log g &Teff&  log Z&  L/HP&  a$_4$(KEPLER)\\
\hline
log g &Teff&  log Z&  L/HP&  $\mu_{cri}$(KEPLER)\\
\hline
log g &Teff&  log Z&  L/HP&  $\sigma$(KEPLER)\\
\hline
....&  ......&  ...&  ...&    ......\\
4.50&  5600.0&  0.0&  2.0&    2.0010\\
4.50&  5600.0&  0.0&  2.0&   -3.3488\\
4.50&  5600.0&  0.0&  2.0&    3.6887\\
4.50&  5600.0&  0.0&  2.0&   -1.3371\\
4.50&  5600.0&  0.0&  2.0&    0.0599\\
4.50&  5600.0&  0.0&  2.0&    0.0096\\
....&  ......&  ...&  ...&    ......\\
\hline
\hline
\end{tabular}
\end{flushleft}
\end{table}

In Fig. 1 we show  the angular distribution of the specific intensity for a spherical model PHOENIX-DRIFT with   T$_{\rm eff}$=2600 K and log g=5.0. We note that the traditional method gives a good adjustement but only up to  $\mu\approx$ 0.12, while the FSM provides a much better fitting for all values of  $\mu$, mainly near the limb. For a PHOENIX-COND 
[5600 K, 4.5] model the situation is similar to the previous model, except that
the value of $\mu$ for which the traditional method gives a good fitting is changed to 0.2, instead of  0.12 (Fig. 2). 

On the other hand, in Fig. 3 we can see the differences between  models with the same log g and effective temperature  [2800 K, 5.0] but computed with different input physics. For more details on the COND and DRIFT models, see Husser et al. (2013) and  Witte et al. (2009), respectively. The respective $\mu_{cri}$ are similar, though the slope of the DRIFT model is larger.

The merit function  is defined as

\begin{eqnarray}
{\sigma^2}= \frac{1}{N-M}\sum_{i=1}^{N} \left( {y_i - Y_i}\right)^2
,\end{eqnarray}

\noindent
where $y_i$ is the model intensity at point $i$, $Y_i$ the fitted function at the same point, $N$ is the number of points, and $M$ is the number of coefficients to be adjusted.  This function is useful for evaluating the quality of the fittings. 
Figure 4 shows the superiority of the new FSM over the traditional one. In this figure   the square root of the merit function for all PHOENIX-COND models is shown for the FSM (upper panel) and for the traditional method (lower panel). The panels are on the same scale to facilitate comparison. 
It is clear that the quality of the FSM fitting is much better than that obtained with the traditional method  for any effective temperature or log g. On average,  the merit function is approximately ten times smaller for the FSM.
This  scenario is similar for all photometric systems examined here and also  for   the monochromatic calculations. 

The effects of the local gravity on the angular distribution of the specific intensities can be seen in Fig. 5 for the same input physics (PHOENIX-COND). As expected, the  $\mu_{cri}$ for  the giant star model is notoriously  larger than for the  main-sequence model at  a fixed effective temperature.   The effects of the effective temperature (for a fixed log g) on the intensity distribution are displayed in Fig. 6, the colder model showing more pronounced slope up to the  $\mu_{cri}$. 

On the other hand, in Fig. 7 we can see the adjustment for the quadratic law (Eq. 2), instead of Eq. 1 (see  Fig. 1 for comparison). As expected, the goodness in this case it is a bit worse than that provided by  Eq. 1, but it is still acceptable. Before adopting either Eq. 1 or  Eq. 2,  the user should check the quality of the adjustment through the respective merit function.

In order to check the accuracy of the FSM, we also performed monochromatic calculations for several models scanning T$_{\rm eff}$ and log g. We  obtained the same results: the FSM is always superior to the traditional method and also describes the regions near the limb  much more accurately. Among  the advantages of the law (Eq. 1) and the application of the FSM, we can list the following: 
a)  it uses a single law which is valid for the whole HR diagram;
b) it is capable of reproducing the specific intensity distribution very well, including the drop-offs; 
c) the flux is conserved within a very small tolerance;
d) it is applicable to different photometric systems  and to
monochromatic values; and
e) it is applicable to different chemical compositions, effective
temperatures, local gravities, and microturbulent velocities.

The present spherical LDC can be used to study the effects of stellar atmosphere spherical models in the diverse environments where they are necessary, such as exoplanetary transits, eclipsing binaries, etc. 
The instruments installed on board the space missions, due to their high photometric accuracy, are especially suitable for trying to detect such effects. We performed calculations for four missions: {{\sc TESS}}, {{\sc KEPLER}}, {{\sc COROT}}, and {{\sc MOST}}. Additional  calculations for other photometric systems and/or other bi-parametric laws can be performed on request.  
The new coefficients and the condition ${I(\mu)}/{I (1)} \equiv 0.0$ for $\mu < \mu_{cri}$ can easily be adapted to the most commonly used light curve synthesis codes.

Finally,  Table A2 summarizes the kind of data available  at the CDS (Centre de Donn\'ees Astronomiques de Strasbourg)  or directly  from the author. They contains the LDC  $a$, $b$, a$_k$, and the $\mu_{cri}$ as a function of the effective temperature and log g. The merit function is also listed to guide the users when selecting the
more suitable spherical LDC.   Table A1 shows an excerpt from Table 7 to help  guide  the reader.

\begin{acknowledgements} 
        I would like to thank  G. Torres,  J. Irwin, and B. Rufino  for the helpful  comments and  T.-O. Husser and P. Hauschildt  for providing the PHOENIX models.   I also thank the anonymous referee who provided good suggestions for improving the manuscript.   
    The Spanish MEC  (AYA2015-71718-R and ESP2017-87676-C5-2-R) is gratefully acknowledged
    for its support during the development of this work. 
        This research has made use of the SIMBAD database, operated at 
        the CDS, Strasbourg, France, and of NASA's Astrophysics Data System Abstract Service.
\end{acknowledgements}

{}

\clearpage
\begin{table}
        \caption[]{Limb-darkening coefficients for the space mission  photometric system FSM}
        \begin{flushleft}
                \begin{tabular}{lccccccl}                         
                        \hline                         
                        Name    & Source   &  range T$_{\rm eff}$ & range log $g$ & log [A/H] & Vel Turb. & Filter & Fit/equation/model   \\ 
                        \hline   
                        Table2 &{\sc PHOENIX-DRIFT}& 1500 K-3000 K & 2.5-5.5&  0.0  & 2 km/s&{\sc TESS}  & LSM/Eq. 1, FSM\\
                        Table3 &{\sc PHOENIX-COND}& 2300 K-12000 K & 2.5-6.0&  0.0  & 2 km/s&{\sc TESS} & LSM/Eq. 1, FSM\\
                        Table4 &{\sc PHOENIX-DRIFT}& 1500 K-3000 K & 2.5-5.5&  0.0  & 2 km/s&{\sc TESS}  & LSM/Eq. 2, FSM\\
            Table5 &{\sc PHOENIX-COND}& 2300 K-12000 K & 2.5-6.0&  0.0  & 2 km/s&{\sc TESS} & LSM/Eq. 2, FSM\\
            Table6 &{\sc PHOENIX-DRIFT}& 1500 K-3000 K & 2.5-5.5&  0.0  & 2 km/s&{\sc KEPLER}  & LSM/Eq. 1, FSM\\
                        Table7 &{\sc PHOENIX-COND}& 2300 K-12000 K & 2.5-6.0&  0.0  & 2 km/s&{\sc KEPLER} & LSM/Eq. 1, FSM\\
            Table8 &{\sc PHOENIX-DRIFT}& 1500 K-3000 K & 2.5-5.5&  0.0  & 2 km/s&{\sc KEPLER}  & LSM/Eq. 2, FSM\\
            Table9 &{\sc PHOENIX-COND}& 2300 K-12000 K & 2.5-6.0&  0.0  & 2 km/s&{\sc KEPLER} & LSM/Eq. 2, FSM\\                  
                        Table10 &{\sc PHOENIX-DRIFT}& 1500 K-3000 K & 2.5-5.5&  0.0  & 2 km/s&{\sc COROT}  & LSM/Eq. 1, FSM\\
                        Table11 &{\sc PHOENIX-COND}& 2300 K-12000 K & 2.5-6.0&  0.0  & 2 km/s&{\sc COROT} & LSM/Eq. 1, FSM\\
                        Table12 &{\sc PHOENIX-DRIFT}& 1500 K-3000 K & 2.5-5.5&  0.0  & 2 km/s&{\sc COROT}  & LSM/Eq. 2, FSM\\
            Table13 &{\sc PHOENIX-COND}& 2300 K-12000 K & 2.5-6.0&  0.0  & 2 km/s&{\sc COROT} & LSM/Eq. 2, FSM\\
                Table14 &{\sc PHOENIX-DRIFT}& 1500 K-3000 K & 2.5-5.5&  0.0  & 2 km/s&{\sc MOST}  & LSM/Eq. 1,  FSM\\
                        Table15 &{\sc PHOENIX-COND}& 2300 K-12000 K & 2.5-6.0&  0.0  & 2 km/s&{\sc MOST} & LSM/Eq. 1, FSM\\
                        Table16 &{\sc PHOENIX-DRIFT}& 1500 K-3000 K & 2.5-5.5&  0.0  & 2 km/s&{\sc MOST}  & LSM/Eq. 2,  FSM\\
            Table17 &{\sc PHOENIX-COND}& 2300 K-12000 K & 2.5-6.0&  0.0  & 2 km/s&{\sc MOST} & LSM/Eq. 2, FSM\\
                        \hline
                        \hline
                \end{tabular}
        \end{flushleft}
\end{table}

\end{document}